\def\ApJ{{\it Astrophys. J.} }

\def\ApJS{{\it Astrophys. J. Suppl.} }

\def\AA{{\it Astron. \& Astroph.} }

\def\MNRAS{{\it Month. Not. Roy. Astr. Soc.} }

\def\simle{\lower 2pt \hbox {$\buildrel < \over {\scriptstyle \sim }$}}
\def\simge{\lower 2pt \hbox {$\buildrel > \over {\scriptstyle \sim }$}}
\documentclass{ws-procs9x6}

\begin{document}

\title{Dark energy - dark matter - and black holes: \\
The music of the universe}

\author{Peter L. Biermann}

\address{Max-Planck Institut for Radioastronomy,  and\\
Department for Physics and Astronomy,\\ University
Bonn, Germany\\ E-mail: plbiermann@mpifr-bonn.mpg.de}


\maketitle

\abstracts{Here we review the recent evidence for dark energy, dark
matter and black holes as components of an expanding universe, for the
vantage point of a non-expert; we speculate on a specific DM
particle.}

\section{Introduction}

In recent years the evidence for the ubiquity of black holes has become
overwhelming, as most galaxies above a certain size appear to have a
massive black hole at their center.  Observations suggest that such black
holes have formed very early in the universe's evolution.

Similarly, we now have convincing evidence, that galaxies are largely
dominated by dark matter, a non-baryonic constituent (Bertone et al.
2005).  

And, finally, in very recent years further evidence has accumulated, that
a dark energy is trying to rip the universe apart.

Here we review very briefly what we know about these strange constituents
of the universe, and speculate on one of them.

\section{What holds the universe together?}

Everyday experience shows that gravity holds us down safely, in our
chair, in our cars, in our trains and airplanes.
The gravitational force of the Earth keeps us down.

\subsection{Throwing a ball}

 From a child throwing a ball to an artifical satellite flying around the
Earth we can describe the orbit with the gravitational pull, the
persistent momentum, and the friction in air as well as the upper
atmosphere.  What is falling down for a ball is curving around the
Earth for a satellite or spaceship.

\subsection{An orbit}

Close to the Earth a satellite flies around in about 90 minutes, and from
the equality {\it centrifugal force = gravitational attraction} we can
calculate the mass of the Earth.  We obtain $5.9723(9) \; 10^{24}$ kg
(see PDG 2004); in this we use the radius of the Earth 6,378.140 km,
already determined in ancient times (although it has to be admitted with
a huge error budget not understood at the time).  Any child can do an
analogous experiment with a pendulum, using a lead-weight and a
suspension via a sharp edge  on two wires.

\subsection{Sun}

The Earth flies around the Sun, at a distance of $1.495,978,706,60(20) \;
10^8$ km, and with a measured velocity of about 30 km/s.  Hence we can
determine the mass of the Sun to $1.988,44(30) \; 10^{30}$ kg.  This is
about 300,000 times the mass of the Earth.  We can also determine the
radius of the Sun to $6.961 \; 10^5$ km.  So light travels from the Sun
to us in 8 minutes, longer than the Sun takes to set.

However, with the Sun we can do a test:  The Sun sounds like a
bell, from its {\it MUSIC} we determine the structure of the Sun.  This
is completey analogous to the quality check of a metal bell, which sounds
only really nice and harmonious if its structure contains no flaws, and
is all homogeneous.  Soundwaves run through a bell, and cause waves to
propagate though the surrounding air.  Similarly soundwaves run through
the Sun, also right through the center.  So we can determine the
structure of the Sun -- and can so test our models for the Sun.

Today we have understood and solved the Solar neutrino problem -- we
understand the Sun -- everything fits; one caveat has to be mentioned:
we still do not understand the origin of magnetic fields in the
universe, and the corresponding challenges of the magnetic plasma near the
Sun may provide the avenue to our insight.  The neutrino problem then gave
deep insight into the properties of neutrinos, their finite mass.

\section{Galaxies and clusters of galaxies}

\subsection{The Milky Way}

The Sun flies around the center of our Milky Way, about once every 250
million years, and at a distance about 20,000 light years.

This motion we can determine within two weeks by measuring
foreground stars against background stars using intercontinental radio
interferometry.  The stars in front rush by the stars in the background,
just like trees in the foreground fly by the houses in the background
looking out from a railroad car; this is of course especially noticeable
if the train goes at 300 km/h.

And so we determine the mass of the Milky Way, and find a mass of about $2
\; 10^{11}$ solar masses.  This is the mass approximately inside the
Solar circle.  Please note that such a determination is strictly accurate
only for a spherically symmetric distribution of mass; for a
rotationally symmetric distribution of mass, for instance, like a
flattened disk system, the gravitational force has to be obtained by a
suitable integral from zero to infinity.  This mass therefore
corresponds only approximately to the mass inside 8 kpc.

In fact, from the observation, that the next neighboring galaxy,
Andromeda or M31, approaches us already, pulled from the overall expansion
of the universe, we can determine an even higher mass, by about a factor
of ten; this mass then corresponds to a radial scale at least about a
factor of ten larger than that obtained at the Solar circle.

However, we can also count stars, to try to determine how many stars are
there, and if their mass is able to account for the mass that we get from
the gravitational pull, exerted on the same stars.  In one approximation
this was done first by J. Oort, who already noted that there seemed to be
a problem:  Depending how we count, we miss anything between a factor of 2
and 10 of the mass in stars in the Galaxy.

\subsection{Galaxies and black holes}

We can repeat this exercise using hot gas in early type galaxies, rotation
curves of cold gas in late type galaxies, and we always find, that we
miss stars.

We can also check for black holes in galaxies, and we always find a
central black hole, in galaxies above a certain size or mass.  In our
Galaxy the mass of the central black hole is about $4 \, 10^6$ solar
masses.  In other galaxies the mass of the central black hole can reach
$3 \, 10^9$ solar masses.  The mass of this black hole appears to be
strictly proportional to the mass of the spheroidal stellar component of
the galaxy, with a factor of 1/500.  There are several explanations for
this proportionality, usually matching star formation in the galaxy with
central accretion (see, e.g., the work by Y. Wang).

\subsection{Clusters of Galaxies: Mass}

Galaxy is just the Greek translation of ``Milky Way", derived from the
Greek word for ``Milk"; we can consider then the motion of galaxies in
clusters of galaxies, and again use both the motions as well as the hot
gas to find out how much gravitational pull is necessary to keep the
cluster from flying apart.

A typical cluster may have a thousand galaxes moving around several
Mega-light-years. From such determinations as well as applying
hydrostatic equilibrium models to the X-ray emission of the hot gas we
obtain a mass of about $10^{15}$ solar masses.

\subsection{Gravitational Lenses = Glasses}

Photons deviate in the gravitational field of the Sun from a straight line
path.  Light  is also bent in its path in and around clusters of
galaxies.  The bending also gives directly the total mass.  We find the
same difficulty, only even more pronounced.

\subsection{Clusters of Galaxies Problem}

We can also count galaxies in clusters, and all the mass in their stars.
And again we can use the X-ray emission from the hot gas to get its mass.
The hot gas even dominates over the stars in mass, integrated all over
the cluster. And the problem is once again:  We are missing out on about
90 \% of the total mass.

What holds clusters of galaxies together?

\section{Matter as you and me}

Matter like you and me we call baryonic matter.  The Earth, Sun, Stars,
hot  and cold gases, all the way to the beginning of the Universe are
made up of baryonic matter.

\subsection{Matter Partner Protons}

At the beginning of the Universe -- after about 3 minutes
-- there was a certain fraction of matter in protons and neutrons. Through
meeting in binary encounters pairs of nucleons can be formed, and so on
Helium.  From today's fraction of Helium in mass we can determine the
initial density of nucleons.  Therefore we can conclude that about 5 - 10
times as much baryonic matter has to be there, than we can directly
observe.  Could this be warm gases with about 100,000. K temperature ?

\section{Music of the Big Bang}

The universe was very hot at the beginning.  Soundwaves just like the
waves which carry my words to you ran through the universe.  The hot
radiation from those early days is still visible.   The waves are also
visible, a disturbance of 1 part in 100,000. We observe, February 11,
2003 on the internet, the first three tones of the Universe, measured by
the satellite WMAP, with incredible precision (astro-ph/0302209, Spergel
et al. 2003).

This is analogous again of the nice sound from a well built bell, or a
violin; a Stradivari sounds wonderful, because of the construction as
well as the materials, the special wood, out of which is has been made.

\subsection{Standard Candles}

Stars explode -- why is still contested. For the massive stars this
may be due to the magneto-rotational mechanism, proposed long ago
(1970) by G. Bisnovatyi-Kogan (many papers 2004/2005), based on an earlier
suggestion by N. Kardashev (1964).  

Stars like the Sun develop into a white dwarf.  White dwarfs in a binary
stellar system obtain more mass via tidal forces.  When they get too much
mass, they collapse, and blow up.  This Supernova explosion seems to be
always the same -- and so we have a standard candle to determine
distances in the universe.

For the lamps along a road, that one can observe from some
mountain late in the evening, we can determine for every lamp its
distance from is apparent brightness, assuming no fog nor haze. And so we
find that the universe expands with acceleration, driven by some DARK
ENERGY with repelling force. Initially this expansion was slow, for some
time it is now accelerating. But why is the factor between DARK MATTER
and DARK ENERGY just 3, and just today?

\section{Matter?}

Normal matter -- that you can see directly -- you, and me, stars, dust and
gases, is an insignificant fraction.  Normal matter -- that one sees only
with great difficulty -- probably warm gases, so total normal matter a
fraction of about 0.04. The relative and dimensionless density here is
relative to the critical density of the universe, a density which would
correspond just to a parabolic orbit without dark energy.  

Invisible matter, that we can determine only through its gravitational
force, so {\bf DARK MATTER}, of some totally different nature, about 0.23
in fraction.  

Invisible energy, called {\bf DARK ENERGY}, about 0.73 in fraction.  

The sum of these contributions is unity to the best of our
understanding (the error budget is about 2 percent).  

So the sum of the angles in any large triangle through the universe is
180 degrees -- so in the language of mathematics -- the universe is flat
just like the geometry of a table in a library.  Of course one should
stay away from black holes for this experiment.

\subsection{Dark matter speculation}

As shown in recent work by F. Munyaneza and A. Kusenko with the author (to
be published), recently several papers have appeared that seem to
converge to a possible solution of the dark matter problem:

\begin{itemize}

\item{}  In Kusenko (2004) it has been shown, that a sterile neutrino in
the mass range 2 - 20 keV can explain the kicks pulsars experience at
birth, giving them linear velocities along their rotation axis up to 1000
km/s.

\item{}  In Munyaneza \& Biermann (2005) it has been shown that a Fermion
dark matter particle far out of thermodynamic equilibrium in its
distribution can explain the early growth of black holes, provided the
mass range is  12 - 450 keV.

\item{}  In Abazajian et al. (2001), and in Mapelli \& Ferrara (2005) it
has been shown that the X-ray observations give a constraint for a sterile
neutrino mass, with one range allowed to be $\simle$ 14 keV.

\end{itemize}

All these lines of reasonings can be summarized as the statement that a
sterile neutrino with a small mixing angle in a mass range near 10 keV is
a possible dark matter candidate.  It appears to fit all observations.
Further checks and specific predictions will be presented in a
forthcoming paper by F. Munyanza, A. Kusenko and the present author.

\section{Riddle}

Let us phrase the riddle in a series of questions:

\begin{itemize}
\item{}  What is this repelling force?
\item{}  Why is the geometry flat?  How is this possible, since the
fractions of all components all vary with cosmic time?  And yet in the
sum always give unity exactly?
\item{}  Half the present age of the universe ago dark matter was ten
times as important as dark energy, in double today's age of the universe
dark energy will be twice as important as dark matter -- will we get the
"BIG RIP"?  Everything in the universe will be torn apart, at last even
the atoms...?
\item{}  Are black holes perhaps important in all this?
\item{}  New particles?  Do I have to look for new particles in the
cosmos instead of at (Geneva) CERN or (Chicago) FERMILAB or (San
Jose) SLAC?
\item{}  What is the interaction between dark matter and black holes?
\end{itemize}

\subsection{What we do not know}

\begin{itemize}

\item{}  Age solar system 4.5 billion years (US billion)
\item{}  Age Milky way about 13 billion years
\item{}  Age Universe 13.7 billion years
\item{}  About $10^{-4}$ fraction black holes
\item{}  About 0.004 visible baryonic matter -- Matter like us
\item{}  About 0.036 invisible baryonic matter: warm gases?
\item{}  About 0.23 non-baryonic matter:  What is this?
\item{}  About 0.73 dark energy:  What is this?
\item{}  Future uncertain:  What does it mean and what does it bring?
\end{itemize}

\subsection{The young minds}

\begin{itemize}

\item{}  {\bf We are looking for an answer}

\item{}  What does this all mean?

\item{}  Please help!

\item{}  Where are our next students?

\item{}  Women and men?

\item{}  Who wants to help looking?

\end{itemize}

\section*{Acknowledgement}

First of all I would like to express my appreciation for the comments on
this ms by A. Kusenko and F. Munyaneza, my partners and friends in the DM
work.  The initial push for the latest stage of this work happened in the
inspiring atmosphere at the Aspen Center for Physics, summer 2005; the
ACP is funded by the National Science Foundation.  Continuing work with F.
Munyaneza is funded by the Humboldt Foundation.

P.L. Biermann would like to acknowledge St. Barwick, G. Bisnovatyi-Kogan,
D. Cline, T. En{\ss}lin, P. Frampton, T.W. Jones, R. Juszkiewicz, H. Kang,
P.P. Kronberg, A. Kusenko, N. Langer, S. Moiseenko, F. Munyaneza, J.P.
Ostriker, D. Ryu, Y. Wang, and T. Weiler for intense discussions of these
and related questions.

Special support comes from the European Sokrates / Erasmus
grants in collaboration with East-European Universities and academy
institutes, with partners W. Bed\-narek, L. Gergely, M. Ostrow\-ski, K.
Petrovay, A. Petrusel, M.V. Rusu, and S. Vidrih, and VIHKOS through the
FZ Karlsruhe.  Recent support comes from NATO for a collaboration with S.
Moiseenko and G. Bisnovatyi-Kogan (Moscow).

Work with PLB is being supported through the AUGER theory and membership
grant 05 CU 5PD1/2 via DESY/BMBF (Germany); further support
for the work with PLB has come from the DFG, DAAD, Humboldt Foundation
(all Germany).


\begin{thebibliography}{0}

\bibitem{}  Particle dark matter: evidence, candidates and constraints,
   Bertone, G., Hooper, D, \& Silk, J., {\it Physics Reports}, {\bf 405},
   279 - 390 (2005), hep-ph/0404175

\bibitem{}  Pulsar kicks from neutrino oscillations, Kusenko, A., {\it
   Int. J. of Mod. Phys. D}  {\bf 13}, 2065 - 2084 (2004),
   astro-ph/0409521

\bibitem{}  Fast Growth of supermassive black holes in Galaxies,
   Munyaneza, F.,  \& Biermann, P.L., \AA  {\bf  436}, 805 - 815
   (2005), astro-ph/0403511  

\bibitem{}  Direct detection of warm dark matter in the X-ray,
   Abazajian, K., Fuller, G.M., \& Tucker, W.H., \ApJ {\bf 562}, 593 - 604
   (2001), astro-ph/0106002

\bibitem{}  Background radiation from sterile neutrino decay and
   reionization, Mapelli, M., Ferrara, A., \MNRAS (submitted 2005)
   astro-ph/0508413

\bibitem{} First-Year Wilkinson Microwave Anisotropy Probe (WMAP)
   Observations: Determination of Cosmological Parameters, Spergel, D.N.,
   et al., \ApJS {\bf 148}, 175 - 194 (2003), astro-ph/0302209; and many
   later papers by the WMAP team

\end{thebibliography}
\end{document}